\documentstyle[aps,prl,multicol,fancyheadings,amsbsy,amssymb,amstex]{revtex}
\begin{document}

\draft
\title
{Zero Temperature Phases of the Electron Gas}

\author{G. Ortiz$^a$, M. Harris$^b$, and P. Ballone$^c$}

\address{(a) Theoretical Division, Los Alamos National Laboratory,  
P.O. Box 1663, Los Alamos, NM 87545 }
\address{(b) Max-Planck Institut f\"ur Festk\"orperforschung,
Heisenbergstrasse 1, 70569 Stuttgart, Germany}
\address{(c) Institut f\"ur Festk\"orperforschung 
Forschungszentrum J\"ulich, 52425 J\"ulich, Germany}

\date{\today}
\maketitle
\begin{abstract}
The stability of different phases of the three-dimensional 
non-relativistic electron gas is analyzed using stochastic methods. 
With decreasing density, we observe a {\it continuous} transition 
from the paramagnetic to the ferromagnetic fluid, with an intermediate
stability range ($25\pm 5 \leq r_s\leq 35 \pm 5$) for the {\it 
partially} spin-polarized liquid. The freezing transition into a 
ferromagnetic Wigner crystal occurs at $r_s=65 \pm 10$. We discuss the 
relative stability of different magnetic structures in the solid 
phase, as well as the possibility of disordered phases.
\end{abstract}

\begin{multicols}{2}

\columnseprule 0pt

\narrowtext

Ever since the pioneering work of Wigner and Seitz \cite{ws} on the 
cohesive energy of metals, the calculation of the ground state energy 
of the interacting electron gas (quantum one component plasma, Fermi 
gas or simply jellium) became the object of considerable theoretical 
interest \cite{ql}. Indeed, the electron gas provides the simplest 
model in which non-trivial magnetic structures and electron 
localization can be realized by varying a single parameter, namely the 
average electron density $\rho$.

In the present paper we investigate the relative stability of various 
broken symmetry phases of the non-relativistic three-dimensional 
electron gas, both fluid and solid, using stochastic methods. We 
report on a macroscopic instability of the Fermi gas which involves 
partial spin-polarization (weak ferromagnetism), in such a way that 
the paramagnetic to ferromagnetic (full spin-polarization) transition 
is not first order, as previously assumed, but a continuous one. 
Moreover we find that the transition to a Wigner crystal occurs at a 
significantly larger density than the value commonly accepted 
\cite{ca}, and that near the quantum freezing transition the fcc and 
bcc crystal phases are nearly degenerate.
 
The jellium model consists of $N$ electrons enclosed in a box of 
volume $\Omega$ (periodically repeated in space)
in the presence of a neutralizing background of 
positive charge. Two parameters characterize its zero temperature 
phase diagram, namely the particle density $\rho = N/\Omega$ and the 
spin polarization $\zeta = |N_{\uparrow}-N_{\downarrow}|/N$, where 
$N_{\uparrow (\downarrow)}$ is the number of spin-up(down) electrons 
($N=N_{\uparrow}+N_{\downarrow}$). The system is governed by the 
Hamiltonian (Hartree a.u.)
\begin{equation}
\widehat{H}= \sum_{i=1}^N \frac{{\bf p}^2_i}{2}+\sum_{i<j}^N \frac{1}{
\mid {\bf r}_i - {\bf r}_j \mid} + \Lambda \ ,
\label{hamil}
\end{equation}
where ${\bf r}_i$ and ${\bf p}_i)$ are the position and linear 
momentum of particle $i$, and 
$\Lambda$ is a constant representing the effect of the background. 
Since we are interested in the macroscopic properties of this model 
system, the thermodynamic limit ($N, \Omega \rightarrow \infty$, 
keeping $\rho$ constant) is to be performed in the end by finite size 
extrapolation. 

Elementary scaling arguments indicate that the kinetic energy term in 
(\ref{hamil}) goes as $1/r_s^2$ ($r_s$ is the Wigner-Seitz radius in 
units of the Bohr radius and its relation to the density is $\rho^{-1} 
= \frac{4 \pi}{3} r_s^3$), while the potential energy scales as 
$1/r_s$. Depending on the relative strength between Coulomb and 
kinetic energies, we can characterize three different regimes: the 
weak ($r_s \lesssim 1$), intermediate ($1 \lesssim r_s \lesssim 10$), 
and strong ($r_s \gtrsim 10$) Coulomb coupling regimes. The 
%random-phase approximation (RPA) \cite{ql} provides an accurate 
random-phase approximation \cite{ql} provides an accurate 
description of the weak-coupling regime. The intermediate coupling 
region, of direct interest for density functional calculations, has 
been extensively studied by numerical \cite{ca,ob} and semi-analytic 
methods \cite{stls}. Not surprisingly, the least known regime is the 
strongly correlated one, for which an early quantum Monte Carlo 
calculation \cite{ca} is still the most authoritative study. 

To delve into the strong coupling regime we employ the variational 
(VMC) and diffusion (DMC) quantum Monte Carlo methods. The starting 
point is provided by a variational wave function of the Jastrow type:
\begin{displaymath}
\Psi_T({\cal R}) = J[{\cal R},{\it \Sigma}] \ 
{\rm det}_{\uparrow}[\varphi] \cdot {\rm det}_{\downarrow}[\varphi] 
\ , \ \Psi_T({\cal R}) \in {\rm I\!R} \ ,
\end{displaymath}
where ${\rm det}_{\uparrow(\downarrow)}[\varphi]$ is a spin up(down) 
Slater determinant of one-electron orbitals $\varphi$ that are either 
plane waves (fluid phases) or localized functions (crystal phases). 
In the equation above, ${\cal R} = ({\bf r}_1,\cdots,{\bf r}_N$)
and ${\it \Sigma} =(\sigma_1,\cdots,\sigma_N)$ represent the full set
of positions and spins of the electrons in the system.
Considering only 2-body correlations, the Jastrow factor can 
be written as: $J[{\cal R},{\it \Sigma}]= 
\prod_{i<j}^N \exp [v_{ij}({\mid {\bf r}_i - {\bf r}_j \mid})]$. For 
an infinite system the optimal $v_{ij}(r)$ should decay as $1/r$ at 
large distances to account for the correct plasmon dispersion 
relation. For large but finite systems, however, that is not the 
case and to reproduce the long distance behavior without resorting to 
expensive evaluations of $v'$s by Ewald sums, we consider a finite 
range $v_{ij}$, vanishing outside a sphere of radius $R_t$ tangential 
to the unit cell: $v_{ij}(r)=0$ if $r > R_t$. We 
have verified that this truncation involves an insignificant loss of 
correlation energy at the variational level \cite{Note1}. Inside 
the sphere $v_{ij}$, which is different for like and unlike spin 
electrons, is given by:
\begin{equation}
v_{ij}(r)=\bar{v}_{ij}(r)+A_{ij} \ \exp[-\alpha_{ij} r^2 ] \ ,
\end{equation}
and
\begin{equation}
\bar{v}_{ij}=\left\{ \begin{array} {ll}
\mbox{\Large $\frac{a_{ij}\: r + b_{ij}\: r^2 }{1+c_{ij}\: r + 
d_{ij}\: r^2}$} + s_{ij}  & 
\ , \mbox{$r<R_b$}\\ &  \\
\sum_{k=0}^5 \; f_k \:(r-R_b)^k & \ , \mbox{$r>R_b$}
\end{array}
\right. \ .
\end{equation}
The $\{a_{ij}\}$ are determined by the electron-electron cusp 
condition \cite{cus}, while 
$\{b_{ij}, c_{ij}, d_{ij}, s_{ij}, A_{ij}, \alpha_{ij} \}$ and $R_b$ 
are variational parameters. The $\{f_k\}$ are chosen in such a way
that $\Psi_T$ and its two first derivatives are continuous everywhere. 
The one-electron orbitals used for the crystal phases are exponentials 
of a Pad\'e function:
\begin{equation}
\varphi_{\bf j}({\bf r})=
\exp{\left[ \frac{-k_1 |{\bf r - R_j}|^2}{1+k_2 |{\bf r - R_j}|}
\right]} \ ,
\end{equation}
where the (positive) constants $k_1, k_2$ are also variational 
parameters,
and the fixed vectors $\{ {\bf R_j},\ {\bf j}=1,N \}$ are distributed
on a regular (bcc or fcc) lattice. This analytic form allows 
$\varphi_{\bf j}$ to change from Gaussian close to each ${\bf R_j}$ to 
a decaying exponential far from the localization centers. The fluid 
states are eigenstates of zero total momentum, while the solid ones 
are eigenstates of finite lattice translations with vanishing total 
crystal momentum. Both are eigenstates of the $z$-component of the 
total spin $S_z$ with eigenvalue $\hbar M=\hbar(N_{\uparrow}-
N_{\downarrow})/2$. All the free coefficients are optimized by the 
variance minimization technique introduced in Ref.~\cite{umrigar}. The 
optimized wave functions are used to drive the DMC simulation
\cite{diffus}.

Several features of the phase diagram discussed below depend upon the 
accurate evaluation of tiny energy differences. To provide a basis to 
estimate the reliability of our results, we report here the relevant 
aspects of our computation, including an estimate for the statistical 
and systematic errors. First of all, because of the high quality of 
the wave functions, and the relatively long runs, statistical errors 
are the least important source of uncertainties: for all $r_s$, the 
statistical error is less than $1$ \% of the correlation energy. 
Moreover, again because of the high variational freedom in the wave 
function, the energy gain in going from VMC to DMC is relatively 
small. As expected, the difference $\delta E= E_{tot}^{VMC}-
E_{tot}^{DMC}$ is largest at low $r_s$ \cite{gainDMC}, because the 
$n$-body contributions not included in our Jastrow function become 
more important at high density \cite{kwon}. Moreover, for each $r_s$ 
we find that 
$\delta E(\zeta)$ is larger for $\zeta=0$ than for $\zeta=1$, and, in 
between, the $\zeta$ dependence of $\delta E$ is well represented by 
the simple interpolation: $\delta E(\zeta) =\delta E(0)+(\delta E (1) 
-\delta E (0)) \ \zeta^2$.

For the fluid phase the most important source of uncertainty in our 
calculations is the finite-size extrapolation to the thermodynamic 
limit. We applied the extrapolation scheme proposed in Ref.~\cite{ca}:
\begin{multline}
E_{\infty}(\boldsymbol{\zeta})= \\ E_N(\boldsymbol{\zeta})-\Delta 
t_N(\boldsymbol{\zeta}) + (N/b(\boldsymbol{\zeta})-1/\Delta 
v_N(\boldsymbol{\zeta}))^{-1} \ ,
\end{multline}
where $\boldsymbol{\zeta}=(r_s,\zeta)$, $E_N(\boldsymbol{\zeta})$
is the total energy of the finite periodic system, 
$b$ and $E_{\infty}$ are fitting parameters, $\Delta t_N$ and 
$\Delta v_N$ describe the size dependence of Hartree-Fock
kinetic and exchange energies ($\Delta t_N=
t^{HF}_N-t^{HF}_{\infty}$, $\Delta v_N= v^{HF}_N-v^{HF}_{\infty}$). 
The parameter
$E_{\infty}$ is the extrapolated value of the total energy per 
electron. Following Ref.~\cite{ob} we assume $b(\boldsymbol{\zeta})=
b_0(r_s)+b_1(r_s) \ \zeta^4.$ The parameters $b_0(r_s)$ and $b_1(r_s)$ 
are obtained from the VMC total energies of an extended set of systems 
($N=1062$, $1450$, and $1930$ for $\zeta=0$, $N=531$, $725$, $965$ for 
$\zeta=1$). This form is used to extrapolate all the energies computed 
by VMC and DMC for unpolarized, partially and fully polarized fluid 
systems. For lower densities the extrapolation is less critical but 
still important. We verified that for the crystal phases the finite 
size extrapolation error is comparable to the statistical one if $N > 
500$. Therefore, the results for $N=686$ (bcc) and $N=864$ (fcc) are 
assumed to be equal to the $N \rightarrow \infty$ limit.

Extensive DMC calculations are performed for systems with $N=1062$ 
(paramagnetic fluid), $N=725$  (ferromagnetic fluid), $N=686$ (ferro- 
and antiferro-magnetic bcc crystal), $N=864$ (ferromagnetic fcc 
phase). For $r_s \leq 10$ our results agree, within the error bar, 
with those reported in Ref.~\cite{ob}. Representative values for the
correlation energy extrapolated to the thermodynamic limit are 
reported in Table I, and Fig.~\ref{fig1} shows a phase diagram 
displaying 
the stability range of the paramagnetic fluid, ferromagnetic 
fluid, and bcc (ferromagnetic) Wigner phases. It is apparent that a 
transition from the paramagnetic to a ferromagnetic fluid phase takes 
place around $r_s \approx 25$, while from the fluid to the crystal it 
is at $r_s \approx 65$. We found that the fcc is slightly less stable 
than the bcc structure, and more stable than the fluid phase at 
$r_s=70$. The total energy difference between the two Wigner phases 
($10^{-3}$ eV/electron) is five times smaller than the one between 
fluid and bcc crystal, and is, therefore, at the limit of our 
resolution. 

The close competition of the fluid and different crystal forms for 
$r_s \approx 70$ suggests that a metastable amorphous phase could be 
formed in this region of the phase diagram, because the large number 
of available configurations, and the tiny energy differences could 
make the crystallization kinetics exceedingly slow. Preliminary 
calculations, in which the regular lattice of $\{ {\bf R_j } \}$ 
vectors are displaced by random amounts (up to $25$ \% of the nearest 
neighbor distance), show that the total energy increases only very 
slowly with increasing disorder.

We now turn to the analysis of the spin dependence of the total energy
near the ferromagnetic and freezing transitions. To ease the analysis, 
we perform a detailed investigation at the VMC level, and then test 
the validity of the resulting picture by a few DMC calculations.

Regarding the magnetic phase transition, we compute the total energy 
for several sizes and partial spin polarizations \cite{siz} in the 
range $0.8 \leq r_s \leq 30$. The extrapolated results for $r_s=25$ 
are reported in Fig.~\ref{fig2}. It is clear that, at this density, 
the ground state has partial spin polarization (weak ferromagnet). The 
full set of data is used to fit an interpolation for the total energy 
(See Ref.~\cite{perd}):
\begin{multline}
E_{tot}(\boldsymbol{\zeta})=E^{HF}_{tot}(\boldsymbol{\zeta})+
\epsilon_0(r_s)+\\
(\epsilon_1(r_s)-\epsilon_0(r_s)) \cdot (\Gamma_1(r_s) \ 
\zeta^2+\Gamma_2(r_s) \ \zeta^4 ) \ ,
\end{multline}
where $E^{HF}_{tot}(\boldsymbol{\zeta})$ is the Hartree-Fock energy. 
In this expression $\epsilon_{0(1)}(r_s)$ is the correlation energy of 
the paramagnetic(fully polarized) fluid, and together with 
$\Gamma_1(r_s)$ and $\Gamma_2(r_s)$ are fitted to the functional form
\begin{multline}
{\cal G}(r_s)=-2A(1+\alpha_1 r_s) \\
\ln \left[ 1+
\frac{1}{2A(\beta_1 r_s^{1/2}+\beta_2 r_s+\beta_3 r_s^{3/2}+
\beta_4 r_s^2)}\right] \ ,
\end{multline}
whose motivation, and connection to the exact high-density limit for 
$\epsilon_{0(1)}$ are discussed in Ref.~\cite{perd}. This
expression for the energy is minimized with respect to $\zeta$ at 
fixed $r_s$ to identify the ground state spin polarization. The stable 
$\zeta$ as a function of $r_s$ is shown in the inset of 
Fig.~\ref{fig2}. It is manifest that the magnetic transition is a 
continuous one, in contrast to what is predicted on the basis of 
simpler interpolations, like the (widely used) form proposed by Perdew 
and Zunger \cite{pz}. 

We emphasize that the interpolation used to determine the 
ground state polarization is based on VMC calculations. Nevertheless, 
as mentioned above, the total energy difference between the VMC and 
DMC results is small, and has a smooth and predictable dependence on 
$\zeta$. Adding $\delta E(\zeta)$ to the fitted total 
energies does not change the transition from continuous to 
discontinuous, although it moves it slightly towards higher values of 
$r_s$ (e.g. $\zeta=0.5$ polarization is stable at $r_s=26$ in VMC, 
while it is at $r_s=30$ in DMC). We also remind that our calculations 
rely on the fixed-node (FN) approximation, and, at present, we cannot 
exclude that releasing the nodal constraints could change the nature 
of the magnetic transition. However, recent computations reported in 
Ref.~\cite{kwon} show that the FN error decreases rapidly with 
decreasing density, and is small already at $r_s=20$.

Additional evidence supporting the picture of a magnetic instability 
and a continuous phase transition is provided by the analysis of the 
radial distribution function and structure factor, displaying for $r_s 
\approx 30$  long range spin-spin correlations even for 
nominally unpolarized systems ($N_{\uparrow}=N_{\downarrow}$.) This is 
illustrated in Fig.~\ref{fig3}, where the spin-spin radial 
distribution function ($g_{SS}(r)=2[g_{\uparrow \uparrow}(r)-
g_{\uparrow \downarrow}(r)]$) is plotted for two densities close to 
the instability \cite{consist}. At $r_s=20$, $g_{SS}$ displays the 
expected depletion hole for parallel spins, reflecting the preference 
of the system for spin alternation at short distances. At $r_s=25$, 
instead, the spin correlation is small and positive at short range, 
and oscillating at long range, pointing to the formation of magnetic 
domains with partial spin polarization. This interpretation in terms 
of domains is confirmed by careful analysis of snapshot configurations.

Indirect support to our findings comes also from recent experiments 
performed in doped hexaborides (Ca$_{1-x}$La$_x$B$_6$) \cite{expe}. 
These authors report a weak ferromagnetic phase at low carrier 
concentration ($r_s = 28$) with an ordered moment corresponding to a 
partial spin polarization of about 10 \%. Moreover, the observed 
Curie temperature, which is as high as 600 K, is of the same order of 
magnitude as the Fermi energy indicating that this represents the 
natural energy scale of the spin system.

The magnetic phase diagram near the Wigner phase transition is 
somewhat simpler. For $70 \leq r_s \leq 100$ the exchange energy, 
although small in absolute terms, is still one order of magnitude 
larger than the correlation energy, and therefore it is not surprising 
%that the ferromagnetic bcc structure is lower in energy than the 
that the ferromagnetic bcc structure is more stable than the 
anti-ferromagnetic spin ordering. However, both at the VMC and DMC 
level, the energy difference between these two states is very small 
($\sim 1$ \% of the correlation energy), and only the systematic trend 
$E_{tot}^{\em ferro} < E_{tot}^{\em antiferro}$ for $70 \leq r_s \leq 
100$ gives us some confidence in this result. As pointed out in the 
case of the translational disorder, because of entropy and kinetics 
the spin glass phase could be the most relevant one for low density 
electron systems.

In conclusion, using stochastic methods we have studied the magnetic 
and freezing quantum transitions of the fermionic one component 
plasma. We find that, contrary to the commonly accepted picture, the 
paramagnetic to the ferromagnetic fluid transition is a continuous 
one, occurring over the $25 \leq r_s \leq 35$ density range, which has 
been approached in experiments by doping highly correlated solids 
\cite{expe}. The transition to the bcc Wigner crystal is first order, 
it occurs at $r_s=65 \pm 10$ and joins two fully polarized spin 
states. Where 
available, our results for the transition densities disagree 
substantially from those of a previous study \cite{ca}. We think that 
the disagreement is due to the extrapolation of the finite size 
results to the thermodynamics limit. In our case, the importance of 
this extrapolation has been limited by performing computations for 
systems with up to $2000$ electrons, in this way reducing the 
corresponding uncertainties. Finally, we recall that our analysis is 
based on a selected set of broken symmetry states. There exists the 
possibility of other instabilities such as the inhomogeneous 
spin-density-wave state or superconducting states which we have not 
considered in the present calculation.

G.O. acknowledges support from an Oppenheimer fellowship and thanks 
the Aspen Center for Physics for its hospitality. We are thankful to 
E. Abrahams, G. Bachelet, Z. Fisk, M. Jones, R. O. Jones, K. Sturm
and S. Trugman for illuminating discussions.

\begin{table}
%\vskip 0.5cm

\begin{center}
\begin{tabular}{c|r@{}lr@{}lr@{}lr@{}lr@{}l}
$r_s$    & \multicolumn{2}{c}{$1$}
         & \multicolumn{2}{c}{$10$}
         & \multicolumn{2}{c}{$20$}
         & \multicolumn{2}{c}{$50$}
         & \multicolumn{2}{c}{$100$} \\
\tableline
$\zeta=0$& 1&.531   & 0&.497  & 0&.313& 0&.1531 && \\
         & 1&.53     & 0&.512  &  &     &  &      && \\
         & 1&.62     & 0&.505   & 0&.313 & 0&.155  & 0&.0868 \\
$\zeta=1$& 0&.814   & 0&.283  & 0&.1837& 0&.0968 & 0&.0549 \\
         & 0&.794    & 0&.281   &  &     &  &      && \\
         & 0&.86     & 0&.286   & 0&.1950& 0&.0965 & 0&.0564 \\
\end{tabular}
\end{center}
\caption{
Minus the DMC correlation energies (in eV) extrapolated to the 
thermodynamic limit. For each $\zeta$, the first row 
refers to the present results, while the second and third rows 
correspond to Ref. [4] and Ref. [3], respectively. The statistical
error affects the last decimal digit.}
\end{table}

\begin{figure}
\caption{Total energy difference (eV) times $r_s$ of:
($\mathbf \bullet$) the 
paramagnetic and the ferromagnetic fluids; ($\blacksquare$) the 
ferromagnetic bcc crystal and the ferromagnetic fluid. The statistical 
error bar is comparable to the size of the symbols. The ferromagnetic 
fluid is stable when both symbols are above zero.}
\label{fig1}
\end{figure}

\begin{figure}
\caption{Total energy (VMC) as a function of $\zeta$ for a weak 
ferromagnetic state. The statistical error bar is reported for the
lowest energy point. The inset shows the equilibrium spin polarization 
$r_s$ as a function of $\zeta$. For $\zeta \sim 0 $ and $\zeta \sim 1$ 
the results depend significantly on the interpolation (see text), and 
therefore they have not been reported in the figure.}
\label{fig2}
\end{figure}

\begin{figure}
\caption{Spin-spin correlation function (VMC) near the magnetic 
instability.}
%\caption{Critical behavior of the spin-spin correlation function (VMC)
%near the magnetic instability.} 
\label{fig3}
\end{figure}

\end{multicols}


\begin{references}
\bibitem{gainDMC} We find $\delta E$ is $0.027$ eV at $r_s=1$, and 
$0.0059$ eV at $r_s=10$ with $\zeta=0$.
\bibitem{consist} For consistency with Fig. 2, we report in Fig. 3 the 
results of the VMC calculation. A similar picture is obtained by DMC 
at a slightly lower density.
\end{references}
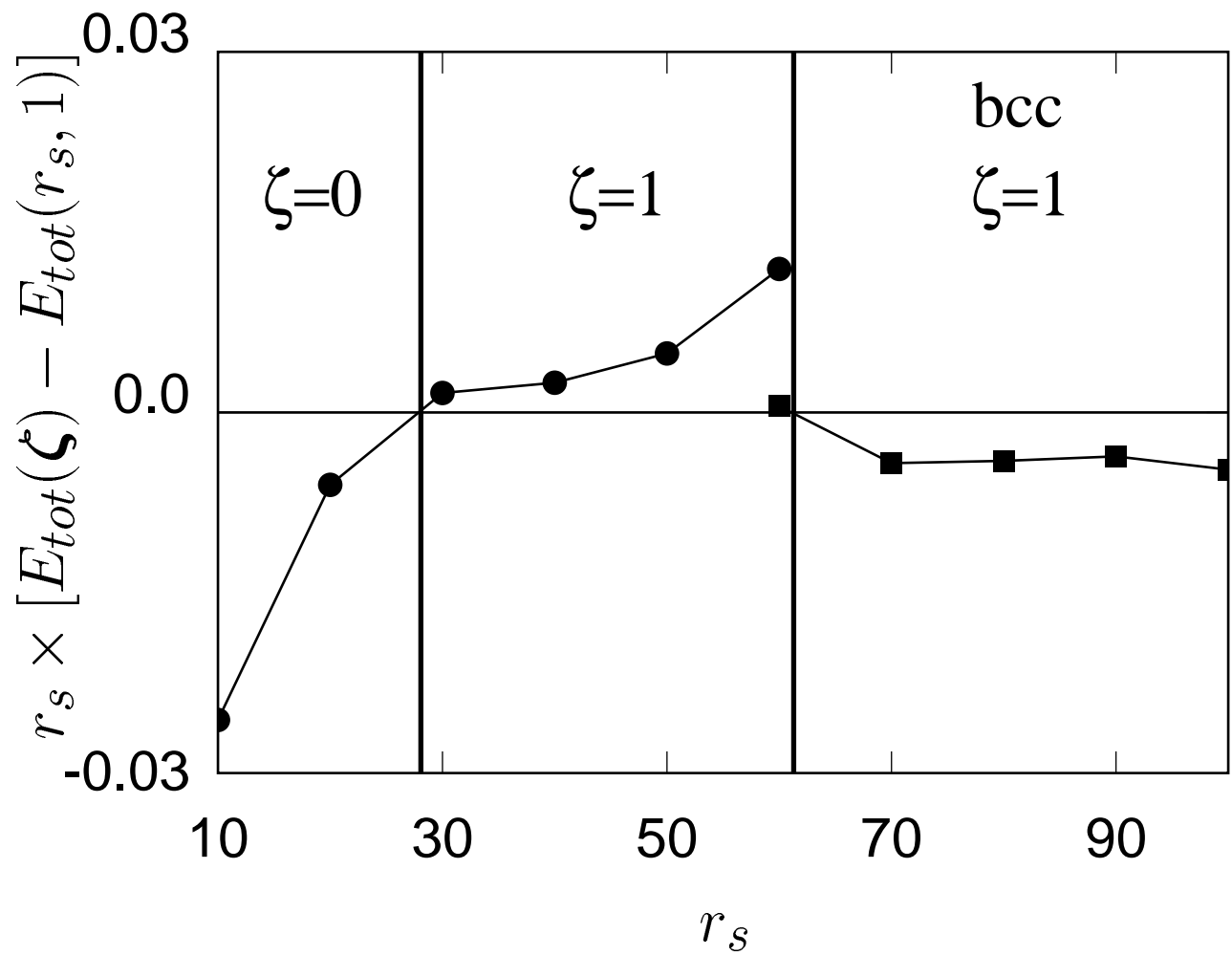
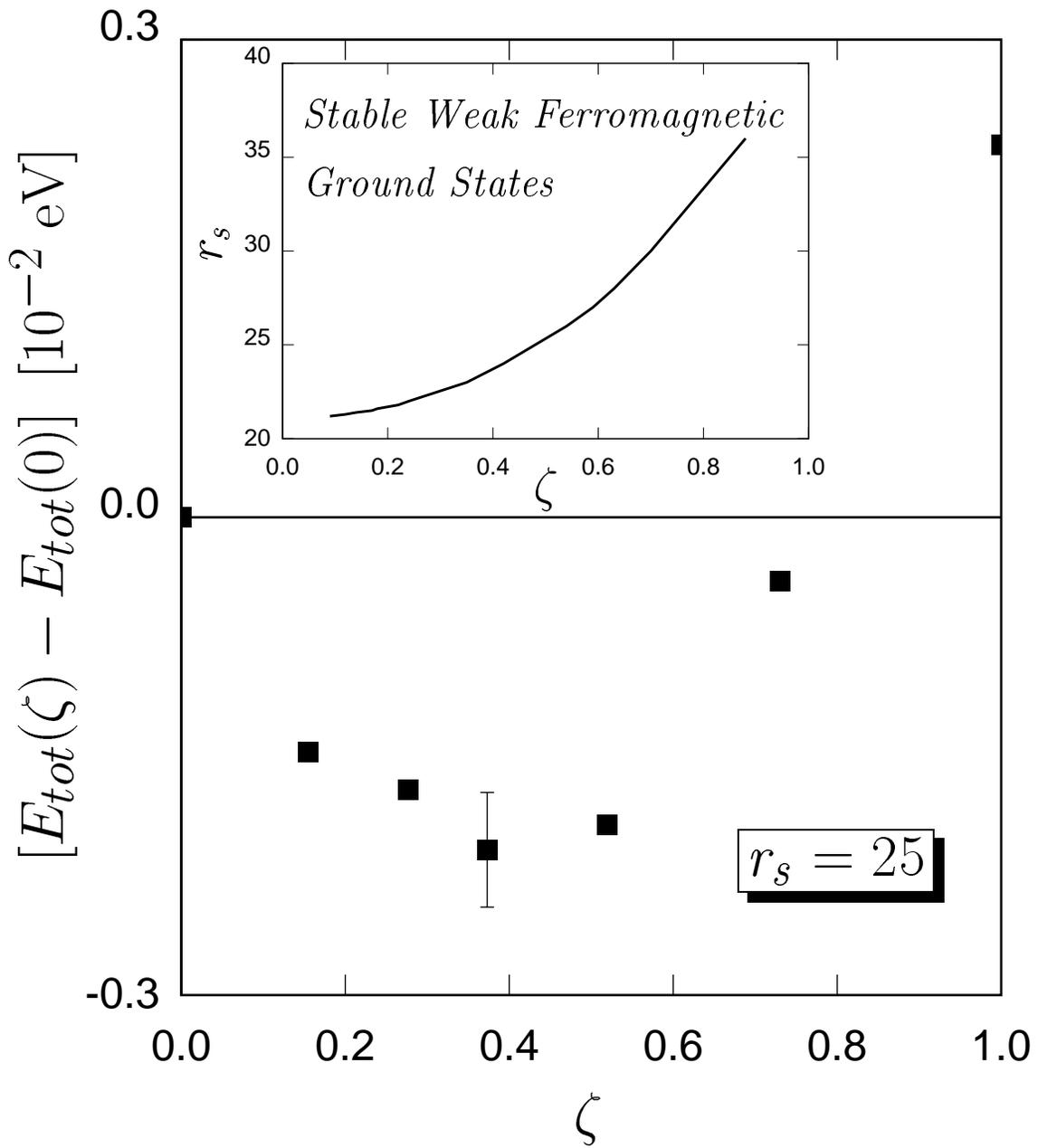
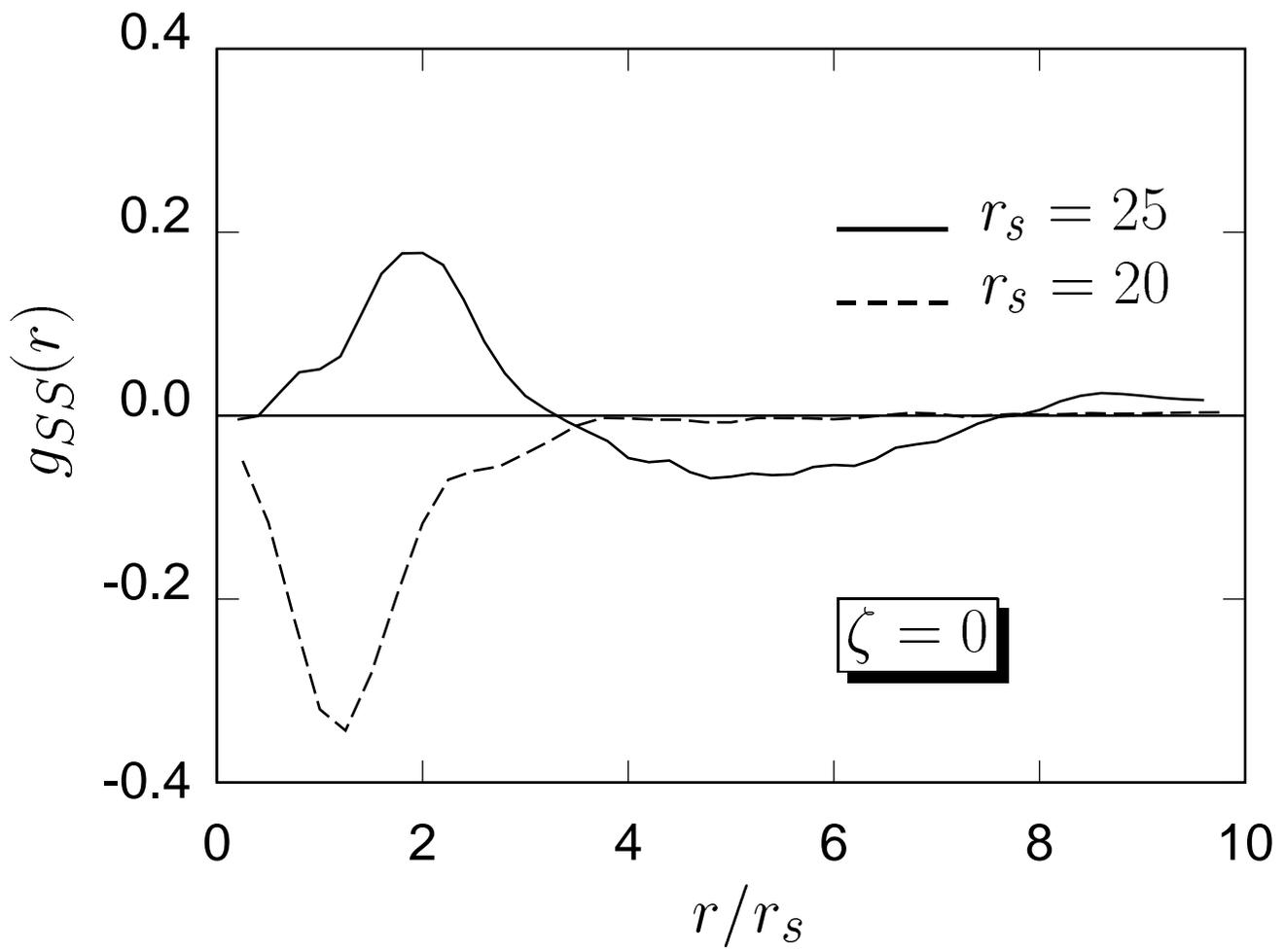
\end{document}